\documentclass[twocolumn,showpacs,prl,preprintnumbers,amsmath,amssymb,aps,floatfix]{revtex4}
\usepackage{graphicx}% Include figure files
\usepackage{dcolumn}% Align table columns on decimal point
\usepackage{bm}
\usepackage[dvips]{epsfig}
\newcommand{\eps}{\varepsilon}
\newcommand{\kk}{{\rm \, K}}
\begin{document}

\title{Memory Effects in the Standard Model for Glasses.}

\newcommand{\e}{{\rm e}}
\newcommand{\veps}{\varepsilon}
\renewcommand{\d}{{\rm d}}
\newcommand{\BEQ}{\begin{equation}}
\newcommand{\EEQ}{\end{equation}}
\newcommand{\BEA}{\begin{eqnarray}}
\newcommand{\EEA}{\end{eqnarray}}

\author{Gerardo Aquino$^{1}$, Armen Allahverdyan$^{2}$ 
and  Theo M. Nieuwenhuizen$^{3}$}
\affiliation{$^1$Max-Planck-Institut f\"ur Physik komplexer
Systeme-N\"othnitzer Str. 38, 01187 Dresden, Germany,}
\affiliation{$^{2}$ Yerevan Pysics Institute, Alikhanian Brothers
street 2, Yerevan, Armenia}
\affiliation{$^{3}$ Institute for Theoretical Physics, University of Amsterdam,
%Valckenierstraat 65,
 1018 XE Amsterdam, The Netherlands}

\begin{abstract}

The standard model of glasses is an ensemble of two-level systems
interacting with a thermal bath.  The general origin of memory effects
in this model is a quasi-stationary but non-equilibrium state of a
single two-level system, which is realized due to a finite-rate cooling
and very slow thermally activated relaxation. We show that single
particle memory effects, such as negativity of the specific heat under
reheating,  vanish  for a sufficiently disordered ensemble. In contrast,
a disordered ensemble displays a collective memory effect [similar to
that described by Kovacs for glassy polymers], where non-equilibrium
features of the ensemble are monitored via a macroscopic observable. An
experimental realization of the effect can be used to further assess the
consistency of the model.

\end{abstract}
\pacs{ 65.60.+a, 61.43.Fs}

\maketitle

{\it Introduction.} Low temperature properties of glassy and amorphous
materials have been an active field of research
 for more than 30 years \cite{anderson}; see
\cite{review} for a review.  Experiments have shown that many
characteristics of amorphous materials, e.g., the temperature dependence
of the specific heat, are universal but different frome those of
crystals. This evidence has captivated much interest in the attempt of
producing a coherent theoretical picture \cite{anderson,review}.  The
two-level system (TLS) model was one of the first models to fit the
experiments. It soon showed to be very successful in describing the
low-temperature properties of glasses, e.g., the linear
temperature dependence of the specific heat, and gained for itself the
definition of ``Standard Model'' for glasses \cite{review}.  With time
this model was improved to account for more features of amorphous solids
\cite{review} and found applications in describing low-temperature
features of proteins \cite{frau}.  A drawback of the model is that there is an excessive
freedom in choosing the distribution of the ensemble parameters. 

{\it Memory effects} arise in the model when due to cooling down to low
temperatures the thermal activation is impeded \cite{huang}. Thus the relaxation time
increases to an extent that for realistic observation times each
TLS is frozen in a non-equilibrium, quasi-stationary state,
which|in contrast to its equilibrium analog|depends on the history of
the relaxation \cite{huang}. Most visible effects of this
non-equilibrium appear during the subsequent reheating, when due to
thermal reactivation the single system specific heat becomes negative
\cite{amjp}.  We shall show however that this single particle memory
effects do not survive the averaging over a sufficiently disordered
ensemble. In contrast, we propose to implement a memory effect, where
{\it due} to the disorder in the ensemble, locally non-equilibrium
features of the system are monitored via a macroscopic (disorder
averaged) observable.  This effect resembles the one implemented by
Kovacs for glassy polymers \cite{kovacs}.  Once the shape of the effect
is sensitive to dynamic (relaxation) and static (disorder) features of
the model, its experimental verification would constitute a way to
further assess the consistency of the model. Analogs of the Kovacs
effect were recently studied for several models of glasses \cite{kov}. 

\begin{figure}[h!]
%[bhb]
%%\vspace{0.2cm}
\includegraphics[width=6cm]{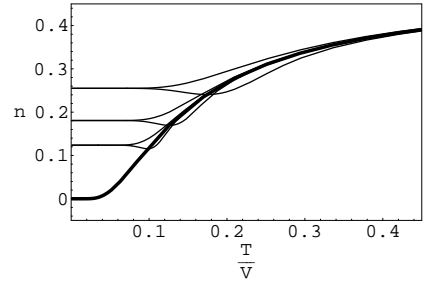}
%\vspace{-0.2cm}
\caption{Upper level occupation $n$ 
vs. dimensionless temperature $T/V$ 
for a single TLS with 
$\eps/V=0.2$. Thick curve: 
%the equilibrium value of $n$
  $n_{eq}$. 
The pairs of normal curves refer to cooling 
(upper curves) and reheating (lower curves) with dimensionless  rate
(from top to bottom) $\sigma=0.01,\,0.001,\, 0.0001$. 
}
\label{fig_00}
%\vspace{0cm}
\end{figure}

{\it The standard model} of glasses amounts to independent
particles, each one moving in an asymmetric double-well potential with
$\varepsilon$ and $V$ being the energy difference between the wells
and the barrier height, respectively \cite{anderson,review}. Each
particle couples to a thermal bath. The positive variables
$\varepsilon$ and $V$ change from one particle to another, so that 
to become observables the single-particle
characteristics, such as energy or specific heat, should be averaged over
the joint distribution $P(\varepsilon,V)$ of $\varepsilon$ and $V$. The
form of $P(\varepsilon,V)$ is well accounted for in literature 
\cite{review}: 
\BEQ
\label{grum}
P(\varepsilon,V)=p_\varepsilon(\varepsilon;\varepsilon_{\rm max}, 
\varepsilon_{\rm min})\,\,p_V(V;V_{\rm min}, V_{\rm max}), 
\EEQ
where $p_\varepsilon$ and $p_V$ are flat distributions with 
%$\varepsilon_{\rm max}$ and $V_{\rm max} $ ($\varepsilon_{\rm min}$ and
 %$V_{\rm min} $)
%being the maximal (minimal) values of $\varepsilon$ and $V$, respectively.
$\eps_{min}<\eps<\varepsilon_{\rm max}$ and $V_{min}<V<V_{\rm max}$.
%$\varepsilon_{\rm min}$ and $V_{\rm min} $)

There are two regimes in the motion of the single system.  {\it i)} The
thermally activated regime is realized when the bath-particle coupling
is sufficiently large. At each moment of time the particle is then
effectively in one of the wells, making sudden jumps between them. The
classical two-state approach is thus a good description of this regime.  {\it
ii)} For low temperatures and weak particle-bath couplings only the
lowest two energy levels of the quantum double well Hamiltonian are
relevant and the problem reduces to a quantum TLS coupled
to a bath \cite{review}. 
Here we study only the classical regime. 
\begin{figure}[h!]
%[bhb]
%%\vspace{0.2cm}
\includegraphics[height=3.9 cm, width=6.5cm]{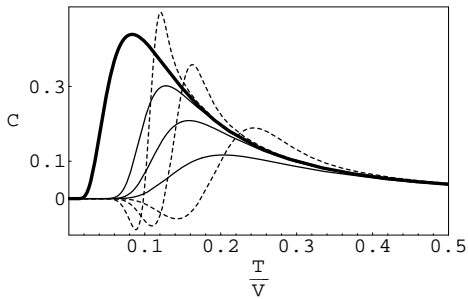}
%\vspace{-0.3cm}
\caption{Specific heat $C$ for a single TLS
vs. dimensionless
temperature $T/V$ and $\veps/V=0.2$. Thick
curve: equilibrium $C$. Normal curves: $C$ during  protocol 
(\ref{2})  with (from left to right) 
$\sigma=0.0001, 0.001, 0.01$. Dashed curves:
the continuation of each previous cooling protocol by heating with the same $\sigma$ (with opposite sign),
%(from left to right) 
%$\sigma=0.0001 , 0.001, 0.01$  
%. In all cases the cooling
%was terminated and changed to heating at temperature
starting at   $T/V=10^{-4}$, i.e. when $n$ has  already relaxed to its zero
temperature value.
}
\label{fig_1}
\end{figure}

Let $n$ and $1-n$ be the
probabilities for the particle to be in the higher and lower well,
respectively. Within the thermally activated dynamics one has:
\BEQ
\label{1}
\dot{n}=\gamma_0 e^{-\beta (V+\varepsilon)}(1-n)-\gamma_0 e^{-\beta V}\,n,
%\dot{n}=\gamma_0 \exp[-\beta (V+\varepsilon)](1-n)-\gamma_0 \exp(-\beta V)\,n,
\EEQ
where
 $e^{-\beta (V+\varepsilon)}$ and $e^{-\beta V}$
%$\exp[-\beta (V+\varepsilon)]$ and $\exp[-\beta V]$
 are the rates of the inter-well motion,
$T=1/\beta$ is the bath temperature, and where $\gamma_0$ is the
attempt frequency. Eq.~(\ref{1}) is solved as
\BEQ
\label{1.1}
n_t=e^{-\frac{t}{\tau}}(n_0-n_{eq})+n_{eq},\quad
\tau=e^{\beta V}/[\gamma_0 (1+e^{-\beta \veps})], 
\EEQ
where $\tau$ is the relaxation time 
and $n_{eq}(\beta)=1/(\e^{\beta \varepsilon}+1)$   the equilibrium
value of $n$ reached for $t\gg\tau$.  At low temperatures the relaxation
time $\tau$ becomes very large, since there is no enough energy for
thermal activation. In this regime  a freezing temperature $T^{*}$ can be
defined (see \cite{review}) below which $n$ is essentially frozen-in at $T=T^*$.  
\begin{figure}[bhb]
%[h!]
\includegraphics[height=3.9cm,width=6.1cm]{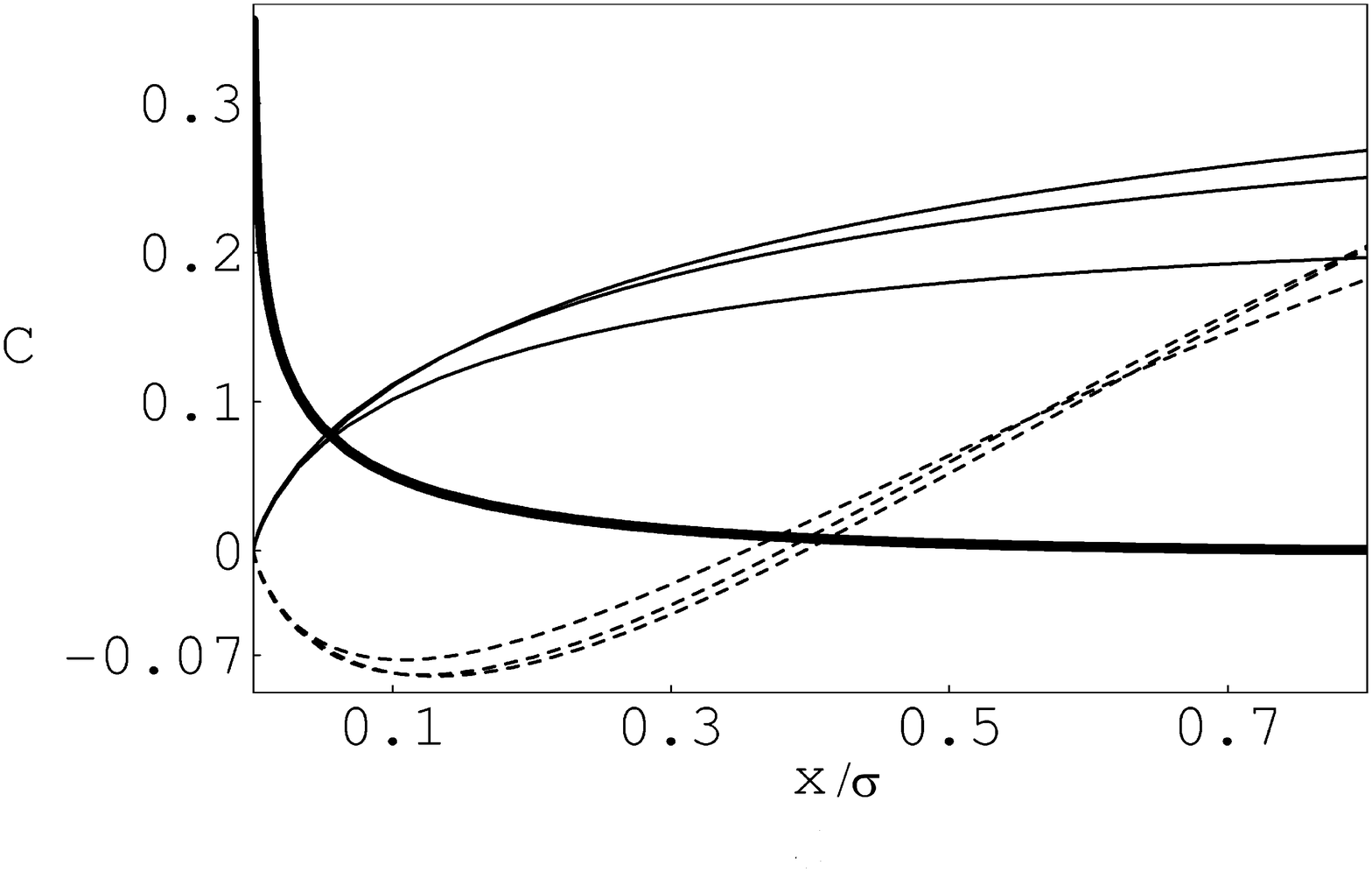}
%\vspace{-0.4cm}
\caption{Specific heat $C$ versus
$x/\sigma={\sigma}^{-1} e^{-V/T}$ for a single TLS
with $\veps/V=0.2$. Thick
curve: equilibrium $C$. Normal (dashed) curves:
$C$ got via cooling (reheating) protocol 
(\ref{2}). 
%from high a 
%temperature at different dimensionless cooling rate.
 From top to bottom:
$\sigma=15\times 10^{-6}, 10^{-4},0.001$. 
}
\label{fig_2}
\end{figure}
{\it Cooling.} Assume that the bath temperature is cooled according to the
following non-linear protocol:
\begin{equation}
\label{2}
\beta_t\equiv 1/T_t=\beta_0 +\omega t
\end{equation}
where $\omega>0$ is the dimensional
cooling rate. This  protocol
is reasonable for low $T$, since it satisfies the third law, not allowing
cooling to $T=0$ in a finite time. Expectedly, for a 
small rate $\omega$ and a high temperature $T_t$, $n$
sticks to its equilibrium value $n_{eq}(\beta_t)$
%$(e^{\beta_t\varepsilon}+1)^{-1}$,
 while for lower $T_t$
there will not be sufficient time to reach this value,
since $\tau$ increases; see Fig.~\ref{fig_00}. 
We rewrite (\ref{1}) as:
\BEQ
%\sigma \d n/ \d x
\sigma \frac{\d n}{\d x}
=(x^{\mu}+1)n(x)-x^{\mu},~~~
\sigma\equiv
\frac{\omega V}{\gamma_0}, ~~~
%\omega V/\gamma_0, ~~~
\mu\equiv\frac{\varepsilon}{V},
%\mu\equiv  \varepsilon/V,
\label{betan}
\EEQ
where the variable $x_t\equiv \exp(-\beta_t V)$ is introduced and $\sigma$  is the dimensionless cooling rate. 
The solution  of (\ref{betan}) is:
\BEA
n( x )&=&
n(x_0)\exp{[\frac{x-x_0}{\sigma}+\frac{x^{\mu+1}-x_0^{\mu+1}}{\sigma(\mu+1)}]}
\nonumber\\
&+& \int_{x}^{x_0} \frac{\d z}{\sigma}\, z^\mu\, 
\exp{[\frac{x-z}{\sigma}+\frac{x^{\mu+1}-z^{\mu+1}}{\sigma (\mu+1)}]},
\label{gensol}
\EEA
where $x_0\equiv\exp(-V/T_0)$. Note from (\ref{gensol}) that the
memory about the initial condition $x_0$ is eliminated for $x_0\gg\sigma$.
If this is satisfied and if 
$\sigma$ is small, the integral in (\ref{gensol}) is approximated as
[$a(z)\equiv z^\mu$, $b(z)\equiv z+z^{\mu+1}/(\mu+1)$]:
\BEQ
\int_{x}^{x_0}\d z\, a(z)\, e^{-\frac{b(z)}{\sigma}}
\simeq\frac{\sigma a(x)}{b'(x)}e^{-b(x)/\sigma}.
\label{do}
\EEQ
%%\begin{gather}
%%\int_{x}^{x_0}\d z\, a(z)\, e^{-\frac{b(z)}{\sigma}}
%%=\int_{x}^{x_0}\d z [a(x)+a'(x)(z-x)+..] \times \nonumber\\
%%e^{-\frac{1}{\sigma}[b(x)+b'(x)(z-x)+\frac{b''(x)}{2}(z-x)^2+..]
%%}\simeq\frac{\sigma a(x)}{b'(x)}e^{-b(x)/\sigma},
%%\label{do}
%%\end{gather}
Eqs.~(\ref{gensol}, \ref{do}) leads to the equilibrium value of $n$:
$n(t)=x^\mu/(1+x^\mu)=n_{eq}(\beta(t))$
%1/(e^{\beta(t) \varepsilon}+1)$.
This, however, holds under neglection of terms 
$a'(x)(z-x)$ and 
%$\frac{1}{2}b''(x)(z-x)^2$
$b''(x)(z-x)^2$
in (\ref{do}). Thus for the validity of the approximation we need:
$a(x)b'(x)\gg \sigma a'(x)$ and $b''(x)\sigma\ll [b'(x)]^2$, which
amounts to $\sigma \mu  x^{\mu-1}\ll (1+x^\mu)^2$, and $\sigma 
\mu\ll x(1+x^\mu)$. For $\mu<1$ and $x<1$ we write the relevant conditions as
\BEQ
\label{karavan}
x\gg \sigma \mu \qquad {\rm or}\qquad T\gg V/[-\ln (\sigma\mu)].
\EEQ
For any finite $\sigma\mu$ this condition breaks down for low
temperatures $x\to 0$. For these temperatures, $x \ll \sigma \mu$, we
obtain a non-equilibrium, stationary (time-independent) value for $n$ by
putting in (\ref{gensol}) $x=0$.  If in addition $x_0\gg\sigma$, 
we put in (\ref{gensol}) $x_0=\infty$ and get for $\sigma^\mu\ll 1$:
\BEQ
n( 0 )=\int_{0}^{\infty} \frac{z^\mu\d z}{\sigma}
e^{-\frac{z}{\sigma}-\frac{z^{\mu+1}}{\sigma (\mu+1)}}=
\sigma^\mu\Gamma(1+\mu)+{o}(\sigma^{\mu}).
\label{be1}
\EEQ
Compared to
% to the equilibrium upper-level probability 
$n_{eq}$,
the non-equilibrium $n$ in (\ref{be1})
depends on the dynamical quantities such as the attempt frequency
$\gamma_0$ and the barrier height $V$: $n$ is smaller for a slower cooling; see Fig.~\ref{fig_00}.

Note that the asymmetry $\mu\not =0$ between the wells is crucial for $n(0)\not =n_{eq}$.
For $\mu \to 0$ 
we get from
 the integral in   Eq. (\ref{be1})
almost equilibrium result:
$n(0)=1/2+ (\mu/4)\ln[\exp(\gamma_E)\sigma/2]
%\frac{1}{2}+ \frac{\mu}{4}\ln\frac{\sigma e^{\gamma_E}}{2}
+{\cal O}(\mu^2)$, where $\gamma_E$ is  Euler's gamma. 

{\it Specific heat}|or the response of the energy
$\eps n$ on the temperature change|provides more visible effects of the memory
on the relaxation history.
Using (\ref{betan}) the equilibrium and the non-equilibrium specific heat
are:
% specific heat is: $C_{eq}=\eps\frac{\d n_{eq}}{\d
% as
%In a non-equilibrium situation it is defined 
\BEA
%C_{eq}&=&\eps\frac{\d n_{eq}}{\d T}=\mu^2(\ln x)^2 x^\mu[1+x^\mu]^{-2}\\
&&C_{eq}=\eps\,\d n_{eq}/\d T=\mu^2(\ln x)^2 x^\mu[1+x^\mu]^{-2}\\
%\nonumber C&=&\veps\frac{\d n}{\d t}\,\frac{\d t}{\d T}=\frac{\mu}{\sigma}\,x\%,(\ln x)^2\,[(x^\mu+1)n(x)-x^\mu].
\nonumber &&C=\veps \, \dot{n}\, (\d t/\d T)= \mu \sigma^{-1} x(\ln
x)^2\,[(x^\mu+1)n(x)-x^\mu].
\EEA
%T}=\mu^2(\ln x)^2 x^\mu[1+x^\mu]^{-2}$.
 Since $C_{eq}$ is zero both for
high and low temperatures, it displays a maximum at some intermediate
temperature; see Fig.~\ref{fig_1}.
 Under cooling from some high
temperatures according to (\ref{2}), the specific heat $C_{c}$ shows
signs of freezing: it is smaller than $C_{eq}$, saturates quicker to
zero, and has a smaller maximum. Let us
now terminate the cooling at some temperature $T_l$ which is low enough
so that the energy $\veps n$ relaxed to its stationary value
(\ref{be1}).  Now heat up the bath using the same protocol (\ref{2})
with $T_0=T_l$ and $\omega<0$, and the same dimensionless rate $|\sigma|$.
%The equilibrium specific heat 
%$C_{eq}$ is the same for cooling and reheating.

In contrast, the specific heat  under heating $C_h$ is
seen to be negative for sufficiently small temperatures \cite{amjp}.
This is related to the decrease of the upper-level occupation $n$ under reheating; see Fig.~\ref{fig_00}.  Moreover,
$C_h\approx -C_c$ at these temperatures: the system keeps memory of
the cooling stage and still decreases its energy after thermal
reactivation. Once $C_h$ reaches its negative minimum, it quickly
increases to the positive maximal value that can be larger than the
maximum of $C_{eq}$: the reheating can bring in more thermal instability;
see Fig.~\ref{fig_1}.  For higher temperatures both
$C_c$ and $C_h$ tend to $C_{eq}$. 
\begin{figure}
%[tht]
%%\vspace{-0.4cm}
%\includegraphics[height=4.75 cm,width=7.25cm]{figg4.eps}
\includegraphics[height=3.6 cm,width=5.9 cm]{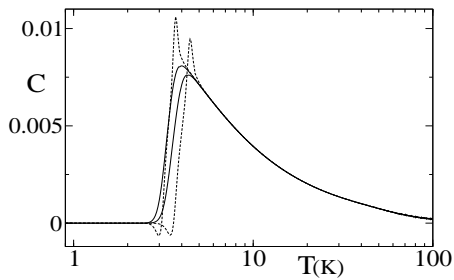}
%\vspace{-0.2cm}
\caption{
Average specific heat $C$ versus time $t$ in cooling from $T_0 = 100K$
(normal curve) to $T_l=0.9 K$ and reheating (dashed curve), with two
different cooling rates: $\omega=10^{-4}(K s)^{-1}$ (curves on the left)
and $\omega=10^{-3}(K s)^{-1}$ (curves on the right). 
The probability of the disorder is given by (\ref{grum}) 
%$P(\eps,V)$ as in(\ref{grum}) 
with $0 \, K < \eps < 5 \, K $ and
$100\kk < V < 400 \kk $. $\gamma_0 = 10^{12}{\rm s}^{-1}$, as for
experiments by A. Nittke {\it et al.} in \cite{review}.}
\label{fig_3}
\end{figure}
\begin{figure}[bhb]
\vspace{0.2cm}
\includegraphics[height=4.4 cm,width=7.0cm]{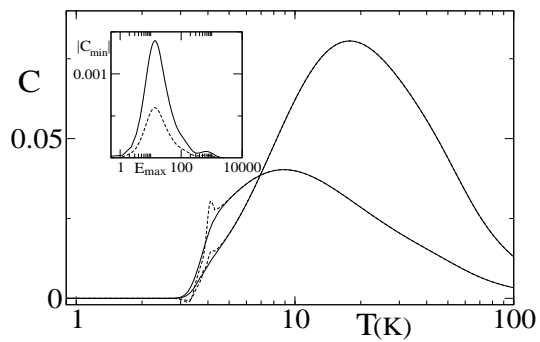}
%\vspace{0.05cm}
\caption{The same protocol  as in Fig.~\ref{fig_3}, but 
with stronger disorder:  $0\kk <\eps< 20\kk$  and $0\kk<\eps< 40 \kk$ (larger curve); 
$\omega=10^{-3} (K s)^{-1}$. In the insert the dependence of 
the minimum negative specific heat as a function of the width of the energy
distribution is displayed for $100 \kk<V<400 \kk$ and $100\kk <V<800 \kk$
(dashed line). $E_{min}=0$.
}
\label{fig_4}
%\vspace{-0.7cm}
\end{figure}
The negativity of $C_h$ shows that the quasi-stationary state of the
TLS cannot be viewed as effective equilibrium, as far as
the reheating is concerned. In order to make the meaning of this result  more clear,
 we note that in the
slow limit, where one decreases $|\sigma|$ and simultaneously increases
the time  $T$ remains constant, we expect
convergence to  equilibrium. Indeed, the temperature region where
$C_h$ is negative, shrinks to zero as $\sim\ln (1/|\sigma|)$ [see
(\ref{karavan})], but the magnitude of the negative   minimal value of
$C_h$ in this region does not depend much on $\sigma$. This is seen upon
plotting $C_h$ versus $x/\sigma=\sigma^{-1} \exp(-V/T)$; see Fig.~\ref{fig_2}.
 However the negativity of $C_h$, and the very difference between $C_h$ and $C_c$, is sensitive to the values
$\varepsilon$ and $V$ of the single-system motion.
%  This, in particular,
%concerns the structure of the scaling variable
%$x/\sigma= \exp(-V/T)/\sigma$ in Fig.~\ref{fig_2}, and the
%position of the  region for negative $C_h$
 
Thus, upon averaging over
the disorder |as given by (\ref{grum}) with experimental values for the
parameters of $P(\varepsilon,V)$| the single-system memory effects
gradually disappear; see Figs.~\ref{fig_3}, \ref{fig_4}.  Even though
each TLS remains non-equilibrium, $C_h$ tends to $C_c$
eliminating the difference between  cooling and heating.
% For stronger
%disorder, $C_h$ and $C_c$ get closer to each other; see
%Figs.~\ref{fig_3}, \ref{fig_4}. 
The same holds for the energy $\eps n$.

We shall now discuss another method for displaying this
non-equilibrium feature. In contrast to the above features which are
essentially single-system and tend to disappear in the presence of disorder, 
the new method is {\it based} on
the presence of an ensemble. 

{\it Temperature shift protocol.} Motivated by Kovacs experiment \cite{kovacs}, we
perform the following protocol:

{\bf 1.} Consider an ensemble of non-interacting TLSs
characterized by a distribution $P(\veps,V)$.  The ensemble is
equilibrated at a given high temperature $T_0$.  
%%: the upperlevel probability of each ensemble member is given by $n_{eq}
%%=\frac{1}{\e^{\beta_0 \varepsilon}+1}$. 

{\bf 2.} Between times $t=0$ and $t_c$ the bath is cooled down following  (\ref{2}).  The cooling is terminated at a
low temperature $T_l$ so that the ensemble averaged energy
$\langle\eps n\rangle\equiv\int \eps\, \d\veps\,\d V\,P(\veps,V)
n(\veps,V)$ reached a stationary value. This determines the  time $t_c$.
Note that $\langle\eps n\rangle$ is observable in experiments using, e.g.,
heat release measurements \cite{review}.
Now $\langle \eps n\rangle$ equals its equilibrium value:
\BEQ
\label{go}
\langle\eps \,n \rangle|_{t=t_c} =\left\langle\eps \,[\,e^{\beta_f\veps}+1\,]^{-1}
\right\rangle\equiv
\langle \eps \,n_{eq}\rangle.
\EEQ
This condition defines the temperature $T_f=1/\beta_f$. If 
most of the TLSs  in the ensemble happen to be described at $t=t_c$ 
by a single temperature, then this temperature will be close to $T_f$ by 
definition. $T_f$ turns out to be of the same order of  the average freezing
temperature $\langle T^*\rangle$.
%averaged over the ensemble.

{\bf 3.} We want to monitor
% via a macroscopic (i.e., ensemble-averaged) quantity 
to what extent the state of the ensemble
 at $t=t_c$ is really close to some internal equilibrium. 
To this end, the bath temperature is suddenly switched to $T_f$, and the
resulting evolution of $\langle\eps n \rangle$ is monitored. 
Due to the sudden switching, the evolution 
is obtained averaging  Eq.~(\ref{1.1})
at  the bath temperature  $T_f$, and with initial state (\ref{go}): 
\BEQ
\label{2.2}\Delta \eps\equiv
\langle\eps n_t\rangle - \langle\eps n_{eq}
\rangle
=\langle\eps e^{-\frac{t-t_c}{\tau}}(n-1/(e^{\beta_f\veps}+1))\rangle.
\EEQ
\begin{figure}
%%\vspace{0.2cm}
\includegraphics[height=4.1cm,width=6.3cm]{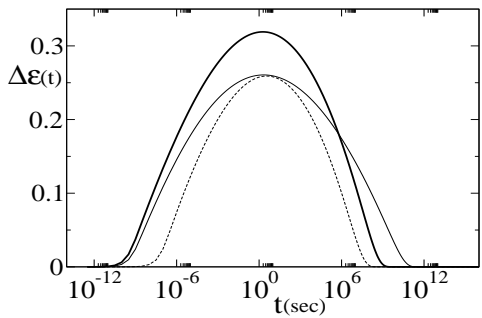}
%\vspace{-0.2cm}
\caption{
$\Delta\eps$ versus time $t$ for the disordered ensemble with
probability distribution given by (\ref{grum}). The cooling protocol (\ref{2}) started
from $T_0=100\kk$ to $T_l=3 K$, with $\omega=10^{-3}\, (\kk {\rm s})^{-1}$.
Thick  curve (stronger disorder): $0 {\rm \,K}
< \eps < 100 {\rm \,K}$, $100 {\rm \,K} < V < 800 {\rm \,K}$, $T_f =
16.6 {\rm \,K} $. Normal curve: $0 {\rm \,K} < \eps < 50 {\rm \,K}$, $100
{\rm \,K} < V < 800 {\rm \,K}$ $T_f = 15.3 {\rm K}$. Dashed curve (weaker disorder): $0
{\rm \,K} < \eps < 50 {\rm \,K}$, $100 {\rm \,K} < V < 400 {\rm \,K}$,
$T_f = 8.8{\rm \,K}$. 
%The thick (normal,dashed) curve represents weak
%(intermediate, strong) disorder.
%the normal curve refers to an intermediate disorder. 
%Note that stronger
%disorder implies a larger temperature $T_f$ and a  higher maximum in the hump
 %of $\Delta\eps(t)$.
 }
\label{fig_8}
%\vspace{-0.4cm}
\end{figure}
It is seen from (\ref{2.2}) that by our construction $\Delta \eps$ 
should be zero both at $t=t_c$ and for a very large
$t-t_c$. It will stay zero for all
times $t>t_c$, if the state of (almost) each  TLS in the
ensemble is described by the same temperature (which need not be equal to
that of the bath). 
Yet another case, where $\Delta \eps$ is constant for $t>t_c$ is when there is no
disorder in the ensemble. Thus, the change of $\Delta \eps$ depends both on 
the disorder and on a non-equilibrium state at $t=t_c$.
The behavior of $\Delta \eps$ for experimentally meaningful parameters
is shown in Fig.~\ref{fig_8}. Since the change of $\Delta\eps(t)$
is finite, a sizable fraction of the ensemble
is at $t=t_c$ far from a local equilibrium.
To gain more understanding, consider the simplest ensemble, which 
is an equal-weight mixture of two TLSs with parameters
$(\veps_1,V_1)$ and $(\veps_2,V_2)$. Eq.~(\ref{2.2}) implies
\BEQ
\nonumber 2 \Delta \eps= 
%\frac{\eps_1}{2}( 
\eps_1 (
e^{-\frac{t-t_c}{\tau_1}}-e^{-\frac{t-t_c}{\tau_2}}
)[\,n(\veps_1,V_1)- 1/(e^{\beta_f\veps_1}+1)],
\EEQ
where $\tau_i=\exp(\beta_f V_i)/[\gamma_0(1+\exp(-\beta_f \veps_i))]$
%$\tau_i=\frac{e^{\beta_f V_i}}{\gamma_0(1+e^{-\beta_f \veps_i})}$
for $i=1,2$ are the relaxation times (see (\ref{1.1})), 
$n(\veps_1,V_1)$ is given by 
(\ref{be1}), and where the temperature $\beta_f$
is defined as in (\ref{go})
 summing over the two TLSs ensemble.
%: $\sum_{i=1,2}\eps_i n(\veps_i,V_i)
%=\sum_{i=1,2}\eps_i [e^{\beta_f \veps_i}+1]^{-1}$.
%%\vspace{1.5 cm}
For the considered simplest ensemble, $\Delta \eps$ is {\it positive}
for $t>t_c$. This is because the slowest system|e.g., system $1$,
if $\tau_1>\tau_2$|has its non-equilibrium upper-level probability 
$n(\veps_1,V_1)$ larger than the
final equilibrium one $[e^{\beta_f \veps_1}+1]^{-1}$. In other words, the slowest
system is further from the equilibrium.
The behavior of $\Delta \eps(t)$ for an experimentally relevant
disorder distributions (\ref{grum}) is displayed in Fig.~\ref{fig_8}.
The fact that $\eps(t)\geq 0$
%|which holds for all cases we were able to check|
implies the same explanation as above: the slow
 TLSs 
%elements of the ensemble
 are further from equilibrium. Two important (and for the present 
effect general) facts seen in
Fig.~\ref{fig_8} is that the stronger disorder leads to {\it i)} larger
value of $T_f$ and {\it ii)} larger maximum of $\Delta\eps$.

 {\it In conclusion}, we studied memory effects 
in the Standard Model for
glasses. This model, besides describing
%in th ensemble of TLSs coupled to the bath. This is known as the
%Standard Model for glasses, since it adequately describes
%low-temperature features of many amorphous materials \cite{review}. In
low-temperature properties of many amorphous materials \cite{review},
%addition, the model
 has important applications in NMR and protein
physics \cite{frau}.  It is known from  previous works
\cite{huang,amjp} that when a single TLS is cooled down to
low temperatures, the relaxation increases due to impeding of the
thermal activation, and the system appears in a quasi-stationary,
non-equilibrium state. In contrast to equilibrium, this state depends on
the detailed features of the relaxation, such as the barrier height or
the cooling rate and upon reheating   manifests itself
%state reflects its
%non-equilibrium features
 via a negative specific heat \cite{amjp}.

We confirmed the latter results by showing that the negative magnitude
of the reheating specific heat is almost insensitive to the decreasing of the
cooling-reheating rate. Next we showed that the single-particle non-equilibrium
(memory) effects disappear for a disordered ensemble. Since only the
latter is experimentally meaningful, one should question whether the
single-particle memory effects can be observed at all. Our main result
is that motivated by Kovacs experiments in \cite{kovacs}, we designed a
protocol which is able to reflect the non-equilibrium features of a
disordered ensemble. The effect is sensitive to the details of the
disorder and, if realized experimentally, it can assess the consistency
of the model. We have also found two universal features of the effect:
{\it i)} it is more visible for a stronger disorder and {\it ii)} its
sign is determined by the fact that slower elements of the ensemble are
further from equilibrium. 

A.E.A. was supported by Volkswagenstiftung and partially  by FOM/NWO.

\end{document}